# 4.4 kV β-Ga$_2$O$_3$ Power MESFETs with Lateral Figure of Merit exceeding 100 MW/cm$^2$

Arkka Bhattacharyya, Shivam Sharma, Fikadu Alema, Praneeth Ranga, Saurav Roy, Carl Peterson, George Seryogin, Andrei Osinsky, Uttam Singisetti and Sriram Krishnamoorthy

*Abstract*—Field-plated (FP) depletion-mode MOVPE-grown β-Ga$_2$O$_3$ lateral MESFETs are realized with superior reverse breakdown voltages and ON currents. A sandwiched SiN$_x$ dielectric field plate design was utilized that prevents etching-related damage in the active region and a deep mesa-etching was used to reduce reverse leakage. The device with $L_{GD}$ = 34.5 µm exhibits an ON current ($I_{DMAX}$) of 56 mA/mm, a high $I_{ON}/I_{OFF}$ ratio > 10$^8$ and a very low reverse leakage until catastrophic breakdown at ~ 4.4kV. The highest measurable $V_{BR}$ recorded was 4.57 kV ($L_{GD}$ = 44.5 µm). An LFOM of 132 MW/cm$^2$ was calculated for a $V_{BR}$ (= $V_{DS}$ – $V_{GS}$) of ~ 4.4kV. The reported results are the first >4kV-class Ga$_2$O$_3$ transistors to surpass the theoretical FOM of Silicon. These are also the highest $I_{DMAX}$ and lowest $R_{ON}$ values achieved simultaneously for any β-Ga$_2$O$_3$ device with $V_{BR}$ > 4kV to date. This work highlights that high breakdown voltages ($V_{BR}$), high lateral figure of merit (LFOM) and high ON currents can be achieved simultaneously in β-Ga$_2$O$_3$ lateral transistors.

## I. INTRODUCTION

Beta-Ga$_2$O$_3$, a unipolar ultra-wide bandgap (UWBG) semiconductor ($E_g$ = 4.6-4.9 eV), has gained increasing importance as a material with tremendous promise to enable power-efficient next generation high voltage power devices. In the last decade of research, β-Ga$_2$O$_3$ material system has witnessed several milestones in bulk and epitaxial single crystal growth, doping, device design, and processing[1]–[4], [4]–[11]. β-Ga$_2$O$_3$-based devices with breakdown voltage up to 8 kV and critical breakdown fields exceeding the theoretical limits of SiC and GaN have been demonstrated[7], [12], [13]. While substantial progress has been made in β-Ga$_2$O$_3$ devices, understanding its material and device physics to take full advantage of its intrinsic properties is still far from mature.

Several field management techniques have been demonstrated in Ga$_2$O$_3$ devices to enhance the average breakdown fields and blocking voltages – the most popular technique being the field-plate (FP) design. But most of these devices suffer from either high reverse leakage that leads to a premature breakdown or low ON currents (and high $R_{ON}$) due to the non-ideal FP process flow involving etching in the gate region [7], [8]. In this letter, we demonstrate over 4 kV-class all-MOVPE-grown β-Ga$_2$O$_3$ lateral MESFETs with a gate FP design using SiN$_x$ field plate/passivation dielectric that achieves high ON currents, low reverse leakage, and LFOM exceeding 100 MW/cm$^2$, simultaneously. We address the critical metrics of $V_{BR}$ (breakdown voltage), $R_{on,sp}$ (specific on-resistance), and ON current ($I_{DMAX}$) at the same time – with significant improvement over the state-of-the-art reports[7], [8], [12], [14], [15].

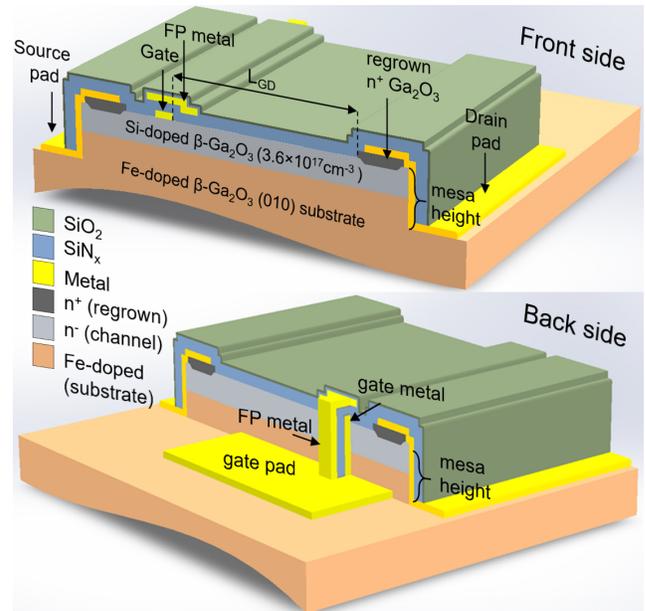

Fig. 1. 3D cross-section schematic of the β-Ga$_2$O$_3$ MESFET showing the FP design. The bottom figure shows the gate FP metal is electrically connected to the gate pad outside the mesa.

## II. DEVICE GROWTH AND FABRICATION

Growth of β-Ga$_2$O$_3$ channel (230 nm thick Si-doped ~3.6×10$^{17}$ cm$^{-3}$) on a Fe-doped (010) bulk substrate was performed by using Agnitron Technology's Agilis 700 MOVPE reactor with TEGa, O$_2$, and silane (SiH$_4$) as precursors and argon as carrier gas. The 10×15 mm$^2$ (010)

This material is based upon work supported by the II-VI foundation Block Gift Program 2020-2021 and the *DoD SBIR Phase I – AF203-CS01 (Contract #: FA864921P0304)*. Work at UB was supported by AFOSR grant FA9550-18-1-0479 (Monitor:Dr. Ali Sayir), NSF ECCS 2019749.

A. Bhattacharyya and P. Ranga are with the Department of Electrical and Computer Engineering, University of Utah, Salt Lake City, Utah, USA 84112. (e-mail: a.bhattacharyya@utah.edu)
S. Sharma and U Singisetti are with the Electrical Engineering Department, University at Buffalo, Buffalo, NY 14260 USA.
F. Alema, G. Seryogin and A. Osinsky are with Agnitron Technology Incorporated, Chanhassen, Minnesota 55317, USA.
S. Roy, C. Peterson and Sriram Krishnamoorthy are with the Materials Department, University of California, Santa Barbara, California, USA 93106. (email: sriramkrishnamoorthy@ucsb.edu)
\* Corresponding author e-mail: a.bhattacharyya@utah.edu

bulk substrate (Novel Crystal Technology, Japan) was cleaned using HF for 30 mins prior to epilayer growth. SF$_6$/Ar ICP-RIE dry etching was utilized for mesa and contact region recessing. The mesa etching was intentionally extended deeper into the substrate, and the total mesa etch height was measured to be ~ 500 nm. The device mesa isolation and the source/drain MOVPE-regrown ohmic contacts fabrication details can be found elsewhere [16]–[20]. Ti/Au/Ni (20 nm/100 nm/30 nm) was evaporated on the regrown n+ contact regions followed by a 450 °C anneal in N$_2$ for 1.5 mins. For the Schottky gate, Ni/Au/Ni (30 nm/100 nm/30 nm) metal stack was evaporated to complete the MESFET structure.

The gate field plate design involved a sandwiched dielectric structure as shown in Figure 1. A 170 nm thick SiN$_x$ film was sandwiched between the gate metal and the FP metal (evaporated Ti (10 nm)/Au (150 nm) /Ni (50 nm)) using sequential metal evaporation and PECVD SiN$_x$ deposition steps. The FP metal was shorted to the gate pad placed away from the device mesa (in the third dimension shown in Figure 1). This FP design avoids dry etching plasma-related damage in the active region. The field plate extension (L$_{FP}$) was 3.2 and 3.5 μm for devices with gate-to-drain length (L$_{GD}$) of 35 and 45 μm. The device mesa was fully passivated using a (50 nm) SiN$_x$/(50 nm) SiO$_2$ bilayer passivation.

## III. RESULTS AND DISCUSSIONS

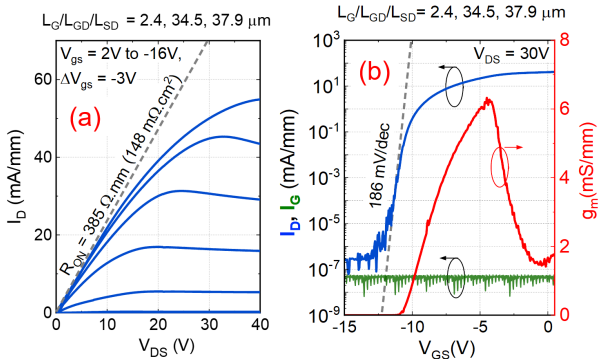

Fig. 2. (a) Output and (b) transfer curves for the β-Ga$_2$O$_3$ MESFETs with L$_{GD}$ = 34.5 μm.

From Hall measurement, the channel charge and mobility were measured to be 5.7×10$^{12}$ cm$^{-2}$ and 95 cm$^2$/Vs respectively (R$_{sh,ch}$ = 11.7 kΩ/□). From transfer length measurements (TLM), the channel R$_{sh,ch}$ was 11.5 kΩ/□ and the total contact resistance to the channel was R$_C$ = 1.4 Ω.mm. The metal to regrown contact layer (R$_{sh,n+}$ ~ 130 Ω/□) specific contact resistance was of the order ~ 10$^{-6}$ Ω·cm$^2$. Fig. 2(a) & 2(b) show the DC output and transfer curves for a MESFET with dimensions L$_{GS}$/L$_G$/L$_{GD}$ = 1.0/2.4/34.5 μm, measured using Keithley 4200 SCS. The reported device dimensions were verified by top-view SEM imaging. The maximum ON current (I$_{DMAX}$) and ON-resistance (R$_{ON}$) measured were ~56 mA/mm and 385 Ω.mm at a gate bias (V$_{GS}$) of 2 V. The contact resistance to the channel, R$_C$, was a negligible part of the total device R$_{ON}$. The devices show sharp pinch-off at a V$_{GS}$ = -13V and low reverse leakage (I$_{ON}$/I$_{OFF}$ > 10$^8$ and negligible gate leakage). A maximum transconductance and sub-threshold swing of 6.2 mS/mm and 186 mV/dec was extracted respectively. The low gate and source-drain leakage indicated minimal surface and bulk-related leakage in these devices.

The breakdown measurements were performed with the wafer submerged in FC-40 Fluorinert dielectric liquid using a Keysight B1505 power device analyzer with N1268A UHV expander. Fig. 3(a) shows the three-terminal breakdown characteristics (at V$_{GS}$ = -20 V) of the MESFET device with L$_{GD}$ of 34.5 um. A breakdown voltage V$_{BR}$ (= V$_{DS}$ – V$_{GS}$) of 4415 V was measured. The device with L$_{GD}$ of 44.5 μm exhibited a V$_{BR}$ of 4567V (not shown). All the devices exhibited very low leakage of 10-100 nA/mm until catastrophic breakdown was observed. The measured reverse leakage currents in Figure 3(a) was limited by the noise floor of the N1268A UHV measurement set-up. Minimizing the reverse leakage was key to achieving the high V$_{BR}$ values and improved L$_{GD}$-V$_{BR}$ linearity. Firstly, the long HF substrate cleaning before the epilayer growth helped in suppressing the parasitic channel at the epilayer/substrate interface that is believed to come from residual Si impurities from the substrate polishing or ambient exposure. As shown from C-V measurements in Figure 3(b), the channel charge profile showed sharp decay near the substrate, indicating the absence of any active parasitic channel. A backside depletion of the channel ~ 50 nm from the substrate was observed which is consistent with the E$_f$ pinning at the Fe trap level (E$_C$ – E$_{Fe}$ ~ 0.8 eV) in the substrate [21]. We hypothesize that the deeper mesa etching was important to eliminate any fringing leakage paths around the device mesa. The low reverse leakage and identical pinch-off voltage values from CV and FET transfer characteristics indicate that these two design steps were very effective in suppressing parasitic channel/charge conduction.

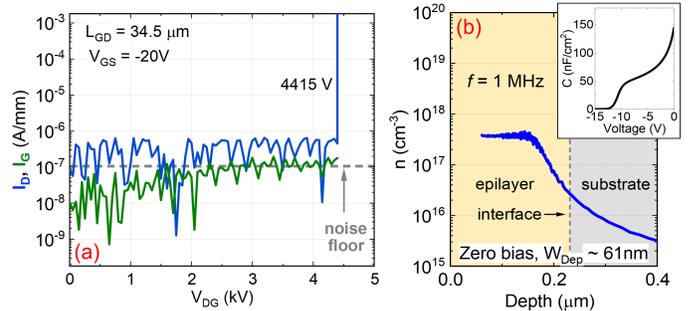

Fig. 3. (a) 3-terminal OFF-state reverse breakdown characteristics of the β-Ga$_2$O$_3$ MESFET with L$_{GD}$ = 34.5 μm. (b) channel charge profile extracted from C-V measurements (inset: capacitance-voltage profile).

Figure 4(a) shows the variation of V$_{BR}$ and I$_{DMAX}$ as a function of L$_{GD}$. The breakdown voltage values exhibit a very linear increase up to L$_{GD}$ = 10 μm (V$_{BR}$ ~ 2.5 kV) and the devices were able to exhibit 2.5 MV/cm average breakdown field (V$_{BR}$/L$_{GD}$). Beyond L$_{GD}$ of 10 μm, the breakdown voltage starts to enter a saturation region and the V$_{BR}$ saturates at around 4.5 kV and does not increase much for L$_{GD}$ of 35 μm to 45 μm. From Sentaurus TCAD simulations, it was estimated that all the devices with L$_{GD}$ ≤ 10 μm had a punch-through (PT) field profile *i.e.* electric field does not go to zero at the drain contact, at their respective breakdown voltages whereas devices with L$_{GD}$ > 10 μm had non-punchthrough (NPT) field profile at breakdown. It can also be seen that the NPT devices (L$_{GD}$ > 10 μm) show larger device-to-device variation in V$_{BR}$ compared to the PT devices (L$_{GD}$ ≤ 10 μm). Figure 4(b) shows the variation of I$_{DMAX}$ with L$_{GD}$ and shows almost a linear change. It is to be noticed that I$_{DMAX}$ values showed very little device-to-device variation unlike the $V_{BR}$ values. This observation indicates that although the contribution from bulk-related leakage paths cannot be completely ruled out, the process variation over the 10 ×15 mm² sample could also lead to the spread in the $V_{BR}$ values. The low device-to-device dispersion in ON currents also indicate that the epi-film conductivity (charge & mobility) is fairly uniform. From TCAD simulations (not shown here), the peak field accumulated in the FP edge in the $SiN_x$ layer and hence dielectric leakage/breakdown could also be limiting the $V_{BR}$ and causing the saturation in $V_{BR}$.

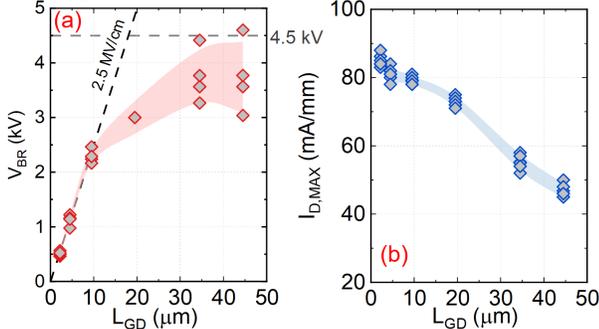

Fig. 4 (a) $V_{BR}$ and (b) $I_{DMAX}$ measured in our β-Ga₂O₃ MESFETs as a function of $L_{GD}$. (shaded region shows device-to-device variation and is a guide to the eye).

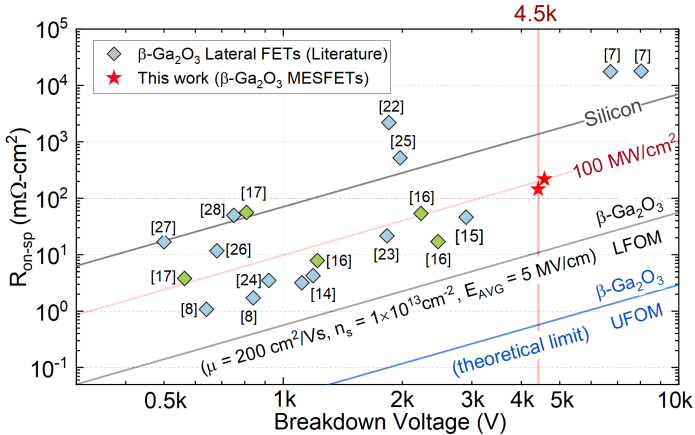

Fig. 5: Differential $R_{on,sp}$-$V_{BR}$ benchmark plot of our β-Ga₂O₃ MESFET with the literature reports [8], [14], [15], [22]–[28].

The lateral figures of merit ($V_{BR}^2/R_{on,sp}$) of these devices were estimated, where $R_{on,sp}$ is $R_{ON}$ normalized to the device active region ($L_{SD}$ +2$L_T$). $L_T$ corresponds to the Transfer Length of the whole ohmic contact (metal to channel) including the regrown layer resistance (2$L_T$ = 0.6 μm) extracted from patterned TLM patterns on the same wafer. An LFOM of 132 MW/cm² was estimated for the device with $L_{GD}$ = 34.5 μm ($V_{BR}$ = 4.4 kV, $R_{on,sp}$ = 148 m Ω·cm² and $L_{SD}$= 37.9 um). The device with $L_{GD}$ = 45 μm exhibited a maximum LFOM of 96 MW/cm² ($V_{BR}$ = 4.57 kV, $R_{on,sp}$ = 219 mΩ·cm² and $L_{SD}$ = 47.9 μm). These values are benchmarked with the existing literarture reports on β-Ga₂O₃ lateral FETs. It can be seen that the devices reported here are the first > 4 kV class β-Ga₂O₃ FET devices to surpass the theoretical unipolar FOM of Silicon. Furthermore, our reported $R_{on,sp}$ are the lowest for any β-Ga₂O₃ FET exceeding a breakdown volatge of 4 kV. The $V_{BR}$-$L_{GD}$ linearity is expected to be further improved by eliminating any parasitic bulk/surface leakage paths, passivation including extreme permittivity materials [29]. The $V_{BR}$-$R_{on,sp}$ trade-off can be further improved by utilizing accumulution channels, improved channel/buffer stack engineering to improve channel mobility in conjunction with minimizing reverse leakage to prevent premature breakdown.

IV. CONCLUSION

In summary, we demonstrate a 4.4 kV class β-Ga₂O₃ lateral MESFET with an LFOM of 132 MW/cm² and ON current of 56 mA/mm – the first > 4 kV-class β-Ga₂O₃ transistor to surpass theoretical UFOM of Silicon. This demonstration shows great promise for MOVPE-grown β-Ga₂O₃ FETs in the low to medium voltage power-device applications.

REFERENCES

[1] M. Higashiwaki and G. H. Jessen, "Guest Editorial: The dawn of gallium oxide microelectronics," *Appl. Phys. Lett.*, vol. 112, no. 6, p. 060401, Feb. 2018.

[2] Pearton, S. J., et al. "A review of Ga₂O₃ materials, processing, and devices." Applied Physics Reviews 5.1 (2018): 011301.

[3] G. Seryogin *et al.*, "MOCVD growth of high purity Ga₂O₃ epitaxial films using trimethylgallium precursor," *Appl. Phys. Lett.*, vol. 117, no. 26, p. 262101, Dec. 2020.

[4] M. Saleh *et al.*, "Electrical and optical properties of Zr doped β- Ga₂O₃ single crystals," *Appl. Phys. Express*, vol. 12, no. 8, p. 085502, Jul. 2019.

[5] J. Jesenovec *et al.*, "Alloyed β-(Al$_x$Ga$_{1-x}$)₂O₃ Bulk Czochralski Single and Polycrystals with High Al Concentration (x = 0.5, 0.33, 0.1)," *arXiv:2201.03673*, Jan. 2022.

[6] K. Sasaki, M. Higashiwaki, A. Kuramata, T. Masui, and S. Yamakoshi, "MBE grown Ga₂O₃ and its power device applications," *Journal of Crystal Growth*, vol. 378, pp. 591–595, 2013.

[7] S. Sharma, K. Zeng, S. Saha, and U. Singisetti, "Field-Plated Lateral Ga₂O₃ MOSFETs With Polymer Passivation and 8.03 kV Breakdown Voltage," *IEEE Electron Device Letters*, vol. 41, no. 6, pp. 836–839, Jun. 2020.

[8] N. K. Kalarickal *et al.*, "β-(Al$_{0.18}$Ga$_{0.82}$)₂O₃/ Ga₂O₃ Double Heterojunction Transistor with Average Field of 5.5 MV/cm," *IEEE Electron Device Letters*, pp. 1–1, 2021.

[9] P. Ranga, A. Bhattacharyya, L. Whittaker-Brooks, M. A. Scarpulla, and S. Krishnamoorthy, "N-type doping of low-pressure chemical vapor deposition grown β-Ga₂O₃ thin films using solid-source germanium," *Journal of Vacuum Science & Technology A*, vol. 39, no. 3, p. 030404, May 2021.

[10] P. Ranga, A. Rishinaramangalam, J. Varley, A. Bhattacharyya, D. Feezell, and S. Krishnamoorthy, "Si-doped β-(Al$_{0.26}$Ga$_{0.74}$)₂O₃ thin films and heterostructures grown by metalorganic vapor-phase epitaxy," *Applied Physics Express*, vol. 12, no. 11, p. 111004, 2019.

[11] S. Roy *et al.*, "In Situ Dielectric Al₂O₃/β- Ga₂O₃ Interfaces Grown Using Metal–Organic Chemical Vapor Deposition," *Advanced Electronic Materials*, vol. 7, no. 11, p. 2100333, 2021.

[12] A. J. Green *et al.*, "3.8-MV/cm Breakdown Strength of MOVPE-Grown Sn-Doped β-Ga₂O₃ MOSFETs," *IEEE Electron Device Letters*, vol. 37, no. 7, pp. 902–905, 2016.

[13] S. Roy, A. Bhattacharyya, P. Ranga, H. Splawn, J. Leach, and S. Krishnamoorthy, "High-k Oxide Field-Plated Vertical (001) β-Ga₂O₃ Schottky Barrier Diode With Baliga's Figure of Merit Over 1 GW/cm2," *IEEE Electron Device Letters*, vol. 42, no. 8, pp. 1140–1143, Aug. 2021.

[14] C. Wang *et al.*, "Demonstration of the p-NiO$_x$/n- Ga₂O₃ Heterojunction Gate FETs and Diodes With BV²/R$_{on,sp}$ Figures of Merit of 0.39 GW/cm² and 1.38 GW/cm²," *IEEE Electron Device Letters*, vol. 42, no. 4, pp. 485–488, Apr. 2021.

[15] Y. Lv *et al.*, "Lateral β-Ga₂O₃ MOSFETs With High Power Figure of Merit of 277 MW/cm2," *IEEE Electron Device Letters*, vol. 41, no. 4, pp. 537–540, Apr. 2020.

[16] A. Bhattacharyya *et al.*, "Multi-kV Class β-Ga₂O₃ MESFETs With a Lateral Figure of Merit Up to 355 MW/cm²," *IEEE Electron Device Letters*, vol. 42, no. 9, pp. 1272–1275, 2021.

[17] A. Bhattacharyya *et al.*, "130 mA mm$^{-1}$ β- Ga$_2$O$_3$ metal semiconductor field effect transistor with low-temperature metalorganic vapor phase epitaxy-regrown ohmic contacts," *Appl. Phys. Express*, vol. 14, no. 7, p. 076502, Jun. 2021.

[18] A. Bhattacharyya, P. Ranga, S. Roy, J. Ogle, L. Whittaker-Brooks, and S. Krishnamoorthy, "Low temperature homoepitaxy of (010) β-Ga$_2$O$_3$ by metalorganic vapor phase epitaxy: Expanding the growth window," *Applied Physics Letters*, vol. 117, no. 14, p. 142102, 2020.

[19] P. Ranga, A. Bhattacharyya, A. Chmielewski, S. Roy, N. Alem, and S. Krishnamoorthy, "Delta-doped β-Ga$_2$O$_3$ films with narrow FWHM grown by metalorganic vapor-phase epitaxy," *Applied Physics Letters*, vol. 117, no. 17, p. 172105, 2020.

[20] P. Ranga *et al.*, "Growth and characterization of metalorganic vapor-phase epitaxy-grown β-(Al$_x$Ga$_{1-x}$)$_2$O$_3$/β-Ga$_2$O$_3$ heterostructure channels," *Appl. Phys. Express*, vol. 14, no. 2, p. 025501, Jan. 2021.

[21] Neal, Adam T., et al. "Donors and deep acceptors in β-Ga$_2$O$_3$." Applied Physics Letters 113.6 (2018): 062101.

[22] K. Zeng, A. Vaidya, and U. Singisetti, "1.85 kV Breakdown Voltage in Lateral Field-Plated Ga$_2$O$_3$ MOSFETs," *IEEE Electron Device Letters*, vol. 39, no. 9, pp. 1385–1388, Sep. 2018.

[23] K. Tetzner *et al.*, "Lateral 1.8 kV β-Ga$_2$O$_3$ MOSFET With 155 MW/cm$^2$ Power Figure of Merit," *IEEE Electron Device Letters*, vol. 40, no. 9, pp. 1503–1506, Sep. 2019.

[24] N. K. Kalarickal *et al.*, "Electrostatic Engineering Using Extreme Permittivity Materials for Ultra-Wide Bandgap Semiconductor Transistors," *IEEE Transactions on Electron Devices*, vol. 68, no. 1, pp. 29–35, Jan. 2021.

[25] K. Zeng, A. Vaidya, and U. Singisetti, "A field-plated Ga$_2$O$_3$ MOSFET with near 2-kV breakdown voltage and 520 mΩ.cm$^2$," *Appl. Phys. Express*, vol. 12, no. 8, p. 081003, Jul. 2019.

[26] Y. Lv *et al.*, "Source-Field-Plated β-Ga$_2$O$_3$ MOSFET With Record Power Figure of Merit of 50.4 MW/cm$^2$," *IEEE Electron Device Letters*, vol. 40, no. 1, pp. 83–86, Jan. 2019.

[27] K. D. Chabak *et al.*, "Recessed-Gate Enhancement-Mode β-Ga$_2$O$_3$ MOSFETs," *IEEE Electron Device Letters*, vol. 39, no. 1, pp. 67–70, Jan. 2018.

[28] M. H. Wong, K. Sasaki, A. Kuramata, S. Yamakoshi, and M. Higashiwaki, "Field-Plated Ga$_2$O$_3$ MOSFETs With a Breakdown Voltage of Over 750 V," *IEEE Electron Device Letters*, vol. 37, no. 2, pp. 212–215, Feb. 2016.

[29] M. W. Rahman, N. K. Kalarickal, H. Lee, T. Razzak, and S. Rajan, "Integration of high permittivity BaTiO$_3$ with AlGaN/GaN for near-theoretical breakdown field kV-class transistors," *Appl. Phys. Lett.*, vol. 119, no. 19, p. 193501, Nov. 2021.